\documentclass[aps,twocolumn,pra,showpacs]{revtex4}
\usepackage{graphicx}

\def\be#1{\begin{equation}\label{#1}}
\def\ee{\end{equation}}
\def\bea#1{\begin{eqnarray}\label{#1}}
\def\eea{\end{eqnarray}}
\def\sp{\hspace{.5em}}
\def\Eq#1{Eq.(\ref{#1})}
\def\Fig#1{Fig.(\ref{#1})}
\def\Figab#1#2{Fig.(\ref{#1}#2)}
\def\no{\nonumber \\}
\def\aM{a_M}
\def\adagM{a_M^{\dagger}}
\def\bI{b_{I}}
\def\bdagI{b^{\dagger}_{I}}
\def\bII{b_{II}}
\def\bdagII{b^{\dagger}_{II}}
\def\mbf#1{\mbox{{\boldmath $#1$}}}

\def\ket#1{|#1\rangle}
\def\ketM#1{|#1\rangle_{M}}
\def\ketI#1{|#1\rangle_{I}}
\def\ketII#1{|#1\rangle_{II}}
\def\ketIII#1#2{|#1\rangle_{I}\otimes|#2\rangle_{II}}
\def\vacM{\ketM{0}}
\def\vacR{\ketIII{0}{0}}
\def\vacI{\ketI{0}}

\def\braI#1{_I\langle#1|}

\def\sOmega{\left( 1-e^{-2\pi\Omega} \right)^{-1/2}}
\def\tOmega{\left( 1+e^{-2\pi\Omega} \right)^{-1/2}}
\def\expnOmega{e^{-2\pi\Omega n}}
\def\expmpiOmega{e^{-\pi\Omega}}
\def\expm2piOmega{e^{-2\pi\Omega}}
\def\Sum{\sum_{n=0}^{\infty}}
\def\Sumnm{\sum_{n=0}^{\infty} \sum_{m=0}^{n}}

\def\sp{\hspace{.25em}}
\def\d#1#2{d^{\sp(#1)}_{\Omega,#2\vec{k}_\perp}}
\def\ddagg#1#2{d^{\sp(#1)\dagger}_{\Omega,#2\vec{k}_\perp}}
\def\b#1#2{b^{\sp(#1)}_{\Omega,#2\vec{k}_\perp}}
\def\bdag#1#2{b^{\sp(#1)\dagger}_{\Omega,#2\vec{k}_\perp}}
\def\a#1{a_{#1\vec{k}_\perp,k^3}}
\def\adag#1{a^\dagger_{#1\vec{k}_\perp,k^3}}
\def\omegak{\omega_{\vec{k}}}
\def\kp{\vec{k}_\perp}
\def\bell#1{\ketM{\mbf{\beta}_{#1}}}

\def\ketML#1{|\mbf{#1}\rangle_{M}}
\def\ketIL#1{|\mbf{#1}\rangle_{I}}

\begin{document}
\title{Teleportation in a non-inertial frame}
\author{Paul M. Alsing}\email{alsing@hpcerc.unm.edu}\author{David McMahon}\email{dmmcmah@sandia.gov}
\affiliation{$^\dag$Center for High Performance Computing and\\
$^{\dag\ddag}$Department of Physics and Astronomy \\
University of New Mexico, Albuquerque, NM 87131}
\author{G. J. Milburn}\email{milburn@physics.uq.edu.au}
\affiliation{School of Physical Science,\\
University of Queensland, Brisbane, Australia}

\begin{abstract}
In this work, we describe the process of
teleportation  between Alice in an inertial frame, and Rob who is
in uniform acceleration with respect to Alice. The fidelity of the
teleportation is reduced  due to Davies-Unruh radiation in Rob's frame.
In so far as teleportation is a measure of entanglement, our
results suggest that quantum entanglement is degraded in non-inertial frames.
We discuss this reduction in fidelity for both bosonic and fermionic resources.
\end{abstract}

\date{\today}
 \maketitle

\section{Introduction}
The large and rapidly growing field of quantum information
science  is a vindication of Landauer's insistence that we
recognize the physical basis of information storage, processing
and communication\cite{Landauer}.  Quantum information science is
based on the discovery that there are physical states of a quantum
system which enable tasks that cannot be accomplished in a
classical world. An important example of such a  task is quantum
teleportation\cite{Bennett-tele}.   Teleportation, like most
recent ideas in quantum information science, is based squarely on
the physical properties of non-relativistic quantum systems.

Recognizing that information science must be grounded in our
understanding of the physical world, one is prompted to ask how
relativistic considerations  might impact tasks that rely on
quantum entangled states. There has recently been some interest in
this question for inertial frames. While Lorentz transformations
cannot change the overall quantum entanglement of a bipartite
state\cite{Peres,Alsing},  they can change which properties of the
local systems are entangled. In particular, Gingrich and
Adami\cite{GA2002} showed that under a Lorentz
transformation the initial entanglement of just the spin
degrees of freedom of two spin half particles can
be transferred into an  entanglement  between both the spin and momentum
degrees of freedom.  Physically this means that detectors, which
respond only to spin degrees of freedom,  will see a reduction of
entanglement when they are moving at large uniform velocity.   Put
simply,  the nature of the entanglement resource depends on the
inertial reference frame of the detectors. A similar result  holds
for photons\cite{Gingrich03}.

In this paper however, we wish to consider quantum entanglement  in
non-inertial frames. In order to make the discussion physically
relevant, we concentrate on a particular quantum information task;
quantum teleportation. We will show that the fidelity of
teleportation is compromised when  the  receiver is making
observations in a uniformly accelerated frame. This is quite
distinct from any reduction in fidelity through the Lorentz mixing
of degrees of freedom noted by Gingrich and Adami\cite{GA2002}.
Rather it is direct consequence of the existence of Davies-Unruh
radiation for accelerated observers.  In so far as teleportation
fidelity is an operational measure of quantum entanglement, our
results suggest that quantum entanglement may not be preserved in
non-inertial frames. While the degree of decoherence  is exceedingly small for practical
accelerations, the apparent connection between space time geometry
and quantum entanglement is intriguing.

The outline of this paper is as follows. In Section II we discuss the essential
features of quantum field theory for an accelerated observer in flat
spacetime. We concentrate our discussion on a massless scalar field, which
will be used to model photons, ignoring polarization. In Section III, we recall the
usual flat space teleportation protocol, and explore the degradation of the fidelity
of the teleported state when one of the participants undergoes
uniform acceleration \cite{alsing_milburn_prl}. We also discuss the reduction of fidelity in terms
of entropy, and the lack of information gain experienced by
the accelerated observer. In Section IV, we extend the previous
bosonic results to the case of Dirac particles, and discuss the similarities
and specific differences. We summarize and conclude our results
in Section V. In an appendix, we discuss a somewhat analogous effect for the case of bosons
in terms of the familiar process of optical parametric down conversion.

\section{Uniformly Accelerated Observers}
\subsection{Preliminaries}
\begin{figure}[h]
\centering
\includegraphics[height=2.5in,width=3.5in]{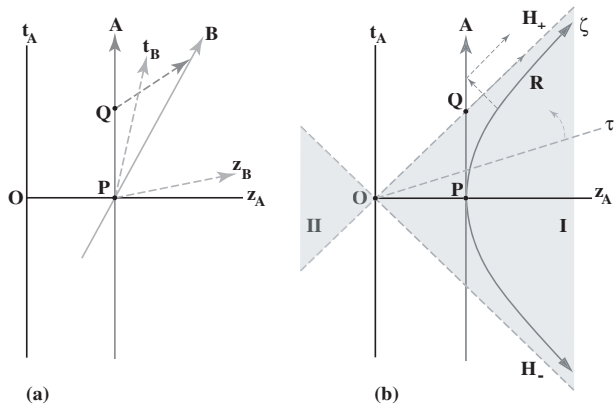}
\caption{(a) Minkowski diagram for the case of Alice (dark gray
arrow) stationary and Bob (light gray arrow) travelling at
constant velocity. Alice and Bob share an entangled Bell state at
the event $P$ (see text). Alice can complete the teleportation
protocol by sending classical signals to Bob at a representative
event $Q$. The entanglement fidelity of a bipartite state $\Psi$ is
unaltered if viewed from either Alice's or Bob's rest frame. (b)
Alice (dark gray arrow) is again stationary, while Rob (dark gray
hyperbola) undergoes constant acceleration. Alice and Rob share a
bipartite entangled Bell state at the common point $P$. The light-like lines
$\mathcal{H}_-$ and $\mathcal{H}_+$ form past and future particle
horizon corresponding to Rob's proper times $\tau = -\infty$ and
$\tau = +\infty$, respectively. At the event $Q$ Alice crosses
$\mathcal{H}_+$ (in her finite proper time $t_A$), and can no
longer communicate with Rob. Rob, however, can still send signals
to Alice across $\mathcal{H}_+$.}\label{accelframe}
\end{figure}
Let Alice be an inertial Minkowski observer with zero velocity, located
at the point $P$ as shown in \Figab{accelframe}{a}.
Another inertial observer Bob is travelling with positive constant velocity $v < c$
in the $z$ direction with respect to Alice,
and their positions are coincident at the point $P$
whereupon they each share one part of an entangled Bell state.
The textbook teleportation protocol \cite{NC} proceeds as usual with Alice sending the
results of her measurement to Bob at the point $Q$, say by photons, so that Bob will
eventually receive them, and be able to rotate his half of the shared entangled state
into the state $\ketM{\psi} = \alpha\ketM{0} + \beta\ketM{1}$ that Alice wishes to teleport
(where the $M$ subscripts denotes a Minkowski state).

The situation is drastically different for the observer Rob who travels with constant acceleration $a$
in the $z$ direction with respect to Alice. Alice's and Rob's position coincide at the point
$P$ where again they instantaneously share an entangled Bell state, of which Rob takes one qubit on his journey.
In Minkowski coordinates Rob's world line takes the form
\be{1}
t_R(\tau) = a^{-1} \sinh{a\tau}, \qquad z_R(\tau) = a^{-1} \cosh{a\tau},
\ee
where $\tau$ is the proper time along the world line.
Rob's trajectory is a hyperbola in Minkowski space bounded by the  light-like asymptotes $\mathcal{H}_-$ and $\mathcal{H}_+$ which
represents Rob's past and future horizons with $\tau = -\infty$ and $\tau = \infty$, respectively.
The shaded region in the right half of the Minkowski plane in \Figab{accelframe}{b}
where Rob is constrained to move is
called the \textit{Right Rindler Wedge} (RRW) and is labelled with the roman numeral $I$. In general,
a point in the RRW can be labelled by the \textit{Rindler coordinates} $(\eta,\zeta)$
which are related to Minkowski coordinates $(t,z)$ by
\be{2}
t = \zeta \sinh{\eta}, \qquad z = \zeta \cosh{\eta},
\ee
where $-\infty < \eta< \infty$ and $0 < \zeta < \infty$. Lines of constant $\zeta$ are hyperbolas
within the RRW and lines of constant $\eta$ are straight lines through the origin. The past horizon
$\mathcal{H}_-$ corresponds to $\zeta = 0, \eta = -\infty$ while the future horizon $\mathcal{H}_+$ corresponds to
$\zeta = 0, \tau = \infty$, both light-like.

With the same coordinate transformation given in \Eq{2}, the
region $-\infty < \zeta < 0$ and $-\infty < \eta < \infty$ is called the \textit{Left Rindler Wedge} (LRW) and
is labeled by the roman numeral $II$. In this region, the lines of constant $\eta$ run in the opposite
sense than in $I$. Region $I$ is causally disconnected from $II$ and no signal from one region can
propagate into the other region. The metric for Minkowski space is given by
\be{3}
ds^2 = d\vec{x}^2_\perp + dz^2 - dt^2 =  d\vec{x}^2_\perp + d\zeta^2 - \zeta^2 d\eta^2
\ee
where $\vec{x}_\perp \equiv (x,y)$.

It is well appreciated now \cite{fulling,unruh,BD,takagi,thooft,pringle} that the quantization
of fields in Minkowski and Rindler coordinates are inequivalent, implying that the
RRW vacuum seen by Rob $\vacI$ is different than the Minkowski vacuum seen by Alice $\vacM$.
The celebrated result of Davies and Unruh \cite{unruh} is that the Minkowski vacuum can be written in terms of
the region $I$ and $II$ states (for a scalar field) as
\be{4}
\vacM = \prod_{\Omega,\kp} \sOmega \Sum e^{-\pi\Omega n} \ketIII{n_{\Omega,\kp}}{n_{\Omega,-\kp}},
\ee
where $\Omega \equiv \omega_R / (a/c)$ with  $\omega_R$ the frequency of a Rindler particle. The Minkowski vacuum as
given by \Eq{4} is a two-mode squeezed state \cite{WallsMilburn} (see appendix \ref{appendix_vacuum_state} )
which for each mode
$(\Omega,\kp)$ has the general form
\be{5}
\vacM \sim \frac{1}{\cosh r} \Sum \tanh^nr \, \ketIII{n}{n},
\ee
with
\be{5b}
\cosh r = \sOmega, \; \sinh r = \expmpiOmega \, \sOmega.
\ee
Note that $\vacM$ can be written as $S(r)\ket{0}_R \equiv S(r)\,\vacR$ where the two-mode squeezing
operator is given by
$S(r) \equiv \exp[r\,(b^\dagger_I b^\dagger_{II} - b_I b_{II})] $ \cite{WallsMilburn}.
The evolution of a Minkowski state vector is effected by the unitary operator
$e^{-i H_M t}$ where for a single mode (notationally ignoring transverse momentum degrees
of freedom) $H_M = \hbar\omega_M \adagM \aM$.
For Rindler states the evolution proceeds via $e^{-i H_R \tau}$ where
\bea{6}
H_R &\equiv& H_I - H_{II} \\
\label{6b}
H_I &=& \hbar\omega_R \bdagI \bI, \qquad H_{II} =\hbar\omega_R \bdagII \bII.
\eea
The minus sign in \Eq{6b} stems from the sense of time essentially flowing
"backwards" in region $II$ (i.e. for $a<0$, $\eta(\tau) = a \,\tau$ is
a decreasing function of $\tau$).

Rob, who lives in region $I$ and is causally
disconnected from the LRW, constructs all his observables
solely in terms of $\bI$ and $\bdagI$ operators, which then act upon $\ket{0}_R$.
Therefore, all physical states in the RRW described by Rob are of the form $\ketI{\psi}\otimes\ketII{0}$.
It is in this sense that we can speak of $\ketI{0}$ as the
"vacuum seen by Rob."

Since he is causally disconnected from region $II$, Rob must reduce any Minkowski
density matrix describing both Rindler wedges to one appropriate
to region $I$ only, by tracing out over region $II$.
For a general Minkowski state $\rho^{(M)}$, the
state $\rho^{(I)}$ perceived by Rob in the RRW is given by
\be{6c}
\rho^{(I)} = Tr_{II}\big(\rho^{(M)}\big).
\ee
In particular, Rob
perceives the Minkowksi vacuum as a thermal state,
\be{7}
\rho^{(I)}_{\vacM} \equiv Tr_{II}(\vacM\langle 0|) = \left(
1-\expm2piOmega  \right) \Sum \expnOmega \ketI{n} \langle n |.
\ee
The exponential terms can be written as $\exp(-\hbar\omega_R/k_B
T_U)$ with the \textit{Unruh temperature} $T_U$ is given by
(in units of $k_B=1$)
\be{8}
T_U \equiv \frac{\hbar a}{2\pi c} = \frac{\hbar}{2\pi c\,\zeta_0},
\ee
where $\zeta(\tau) = \zeta_0 = 1/a$ is the constant
Rindler position coordinate of Rob's stationary world line.

\subsection{Relationship between Minkowski and Rindler modes}
In this section we will use  two-photon states of the
electromagnetic field which, for simplicity and without loss
of generality, are modelled by the massless
modes of a scalar field (we ignore polarization).
For the Bell state used as the entangled resource, we must consider Fock states other than
the vacuum state for the Rindler observer. This is easily done by
a consideration of how the creation and anihilation operators
transform. The relationship between the Minkowski and Rindler
modes is given by the Bogoliubov transformation \be{9}
\b{\sigma}{} = \int d^3 k' \left( \alpha_{kk'}^{(\sigma)}
a_{k'} +
                                  \beta_{kk'}^{(\sigma)} a^\dagger_{k'}
                           \right)
\ee
where the notation of \cite{takagi} has been adopted, namely
 $\sigma = (+,-)$ refers to region $I$ and $II$ respectively, $k=(\Omega,\vec{k}_\perp)$ and
 $k' = (\vec{k}_\perp, k^3)$. Modes in Minkowski space are specified by the wave vector
 $\vec{k}\equiv (\vec{k}_\perp, k^3)$ where $\vec{k}_\perp = (k^1,k^2)$ are the components of
 the momentum perpendicular to the direction of Rob's acceleration. The Minkowski frequency,
 for a general particle of mass $m$, is given by $\omegak = \sqrt{m^2 + \vec{k}^2}$.
 These modes arise from the solution of the scalar wave equation in the standard metric (first
 equality in \Eq{3}).
 Modes in Rindler space arise from the solution of the wave equation in the
 Rindler metric, the second equality in \Eq{3}.
 They are specified by a positive energy Rindler frequency $\Omega$ and $\kp$.
 Solution of the wave equation in region $I$ yields modes that have finite
 support in the RRW, and zero outside. Similarly for region $II$.
 The separate Rindler quantizations
 in the RRW and LRW yield a complete othronormal set of modes that are
 appropriate for their respective regions and independent of the opposite region.
 The Bogoliubov transformation relating Minkowski to Rindler modes
 can be put into a more transparent form by introducing
 a third set of modes, the \textit{Unruh modes} $\d{\sigma}{}$ and $\ddagg{\sigma}{}$.
 The Unruh modes arise
 by considering the Fourier transform of the usual Minkowski plane waves
 $[(2\pi)^3 \, 2\omegak]^{-1/2}\exp(\vec{k}\cdot\vec{x} - \omegak t)$ in terms of
 the Rindler proper time $\tau$ \cite{pringle}. These complete, orthonormal set of modes exits
 over all of Minkowski space and can be "patched together" to form  two complete orthonormal
 set of modes that analytically continue the Rindler modes from their region of
 definition $(I,\,II)$ to their opposite, causally disconnected region $(II,\,I)$.
 The physical significance of the Unruh modes is that they diagonalize the generator of
 Lorentz boosts \cite{takagi}, which in Minkowski coordinates is given by
 $$
    M^{\alpha\beta} = \int d^3x (x^{\alpha} T^{0\beta} -x^{\beta} T^{0\alpha} ).
 $$
 The restriction of the generator of boosts in the $z$ direction
 to region $I$ gives the Rindler Hamiltonian $H_R = \left. M^{03}\right|_I$.

The relationship between the Unruh modes and the Minkowski modes is given by \cite{takagi}
\be{10}
\a{} = \sum_\sigma \int_0^\infty d\Omega p_\Omega^{(\sigma)*}(k^3) \d{\sigma}{}.
\ee
which can be inverted to give
\be{11}
\d{\sigma}{} = \int_{-\infty}^\infty dk^3 p_\Omega^{(\sigma)}(k^3) \a{}.
\ee
In the above expression, the functions $p_\Omega^{(\sigma)}(k^3)$ form
a complete orthonormal set and are given by
\be{12}
p_\Omega^{(\sigma)}(k^3) = \frac{e^{i\sigma\Omega y_{\vec{k}}}}{(2\pi\omegak)^{1/2}}  \quad
y_{\vec{k}} \equiv \frac{1}{2}\ln\left( \frac{\omegak + k^3}{\omegak - k^3}  \right)
\ee
which are just phase factors.
Since by \Eq{11} the Unruh annihilation operator is a sum over only
Minkowski annihilation operators, it too annihilates the Minkowski vacuum.
\be{13}
\a{} \vacM = 0, \qquad \d{\sigma}{\pm} \vacM = 0.
\ee
Finally, the Unruh modes are related in a natural way to the Rindler modes
through the following Bogoliubov transformation
\be{14}
\left[
\begin{array}{c}
  \d{+}{} \\
  \ddagg{-}{-}
\end{array}
\right]
=
\left[
\begin{array}{cc}
  \cosh r & -\sinh r \\
  -\sinh r & \cosh r
\end{array}
\right]
\,
\left[
\begin{array}{c}
  \b{+}{} \\
  \bdag{-}{-}
\end{array}
\right]
\ee
with the hyperbolic functions of $r$ related to the Rindler frequency $\Omega$ by \Eq{5b}.
The operators $b^{(+)}$ and $b^{(-)}$ annihilate the RRW vacuum $|0\rangle_+$ and LRW vacuum
$|0\rangle_-$ respectively, and commute with each other.

By \Eq{10} we see that a given Minkowski mode of frequency $\omegak$ is spread
over all positive Rindler frequencies $\Omega$ (peaked about $\Omega\sim\omegak$), as a result of the
Fourier transform relationship between $\a{}$ and $\d{\sigma}{}$.
We now simplify our analysis by considering the effect of teleportation of
the state $\ketM{\psi} = \alpha \ketM{0} + \beta \ketM{1}$ by the Minkowski
observer Alice to a single Rindler mode of the RRW observer Rob. That is, we consider
only the mode $(\Omega,\kp)$ in region $I$ which is distinct from the mode $(\Omega,-\kp)$
in the same region.
From \Eq{4} the vacuum breaks up into products of pairs of Fock states,
each corresponding to a correlated region $I$-region $II$ mode pair,
\be{14.5}
\ketM{0} \sim \cdots  \{ \ketI{n_{\Omega,\kp}}\ketII{n_{\Omega,-\kp}} \}
 \{ \ketI{n_{\Omega,-\kp}}\ketII{n_{\Omega,\kp}}\}  \cdots
\ee
As such, we can consider only the $\sigma = (+)$ contribution in \Eq{10}
(corresponding to the correlated mode pair in the first set of
parentheses in \Eq{14.5}) and drop the unessential phase factors $p_\Omega^{(\sigma)}(k^3)$.
The single Rindler mode component of
the Minkowski vacuum state we are interested in is then
\be{15}
\vacM \to \frac{1}{\cosh r} \Sum \tanh^nr \ketIII{n_{\Omega,\kp}}{n_{\Omega,-\kp}}.
\ee
The relevant Bogoliubov transformation can now be written as
\be{16}
\adag{} \to \ddagg{+}{} = \cosh r \, \bdag{+}{} - \sinh r \, \b{-}{-}.
\ee
From here on we drop all the frequency and momentum subscripts and replace
the labels $\pm$ by $I$ and $II$, keeping
in mind the full definitions in \Eq{15} and \Eq{16}.

\section{Teleportation from a Minkowski observer to a Rindler observer}
\begin{figure}[h]
\centering
\includegraphics[height=2.5in,width=2.5in]{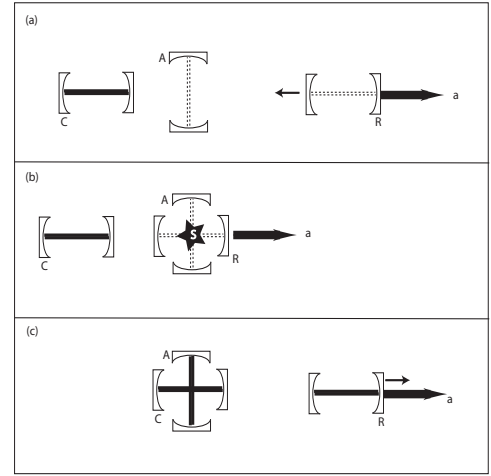}
\caption{Cavities A (Alice, Minkowski), R (Rob, Rindler), and C (client cavity, Minkowski).
(a) For $\tau<0$ Rob's cavity moves with constant acceleration $a$ in the negative $z$-direction.
(b) At $\tau=0$ A and R overlap and a 4-photon entangled state is shared between the two cavities.
(c) At some time $\tau>0$ Alice makes a Bell measurement with the unknown state in
C and her half of the Bell state in A. R moves with constant acceleration in positive $z$-direction.
}
\label{cavities}
\end{figure}

We now discuss how Alice and Rob can come to share an entangled resource for teleportation.
Suppose that Alice and Rob each hold an
optical cavity, at rest in their local frame.  At event P their two frames coincide
when Rob's frame is instantaneously at rest. At this event we suppose that the
two cavities overlap and simultaneously a four photon source excites
a two photon state in each cavity, as depicted in \Fig{cavities}.
We will also assume that, prior to event P, Alice
and Rob ensure that all photons are removed from their cavities.

Suppose that each cavity supports two orthogonal  modes (spatial modes,
we ignore polarization, and model the photons by the massless modes of a scalar field),
with the same frequency,  labelled $A_i,\ R_i$ with $i=1,2$, which
are each excited to a single photon Fock state at event P.  At P the total state held
by Alice and Rob is then the entangled state
\begin{equation}
\label{0.1}
|0\rangle_M\rightarrow |1\rangle_{A_1}|0\rangle_{A_2}|1\rangle_{R_1}|0\rangle_{R_2}
+|0\rangle_{A_1}|1\rangle_{A_2}|0\rangle_{R_1}|1\rangle_{R_2}
\label{four-photon-state}
\end{equation}
where $|1\rangle_{A_i}, |1\rangle_{R_i}$ are single photon excitations of the Minkowski vacuum states in
each of the cavities. Treating these states as single particle excitations of the {\em Minkowski} vacuum is
an approximation. We expect this to be valid so long as the cavities do not move appreciably over the time
taken for the source to excite the modes.  As this time cannot be smaller than the round trip time in the cavity,
we are implicitly assuming the cavities are very small.
 The state in Eq.(\ref{four-photon-state}) encodes the two qubit entangled Bell state
\begin{equation}
\label{0.2}
\ketM{\mbf{\beta}_{00}} \equiv
\frac{1}{\sqrt{2}} \,  \big(\,|{\bf 0}  \rangle_M|{\bf 0}\rangle_M+|{\bf1}\rangle_M|{\bf 1}\rangle_M \,\big)
\end{equation}
where the first qubit in each term refers to cavity A, the second qubit refers to cavity R
and the logical states $|{\bf 0}\rangle_M,|{\bf 1}\rangle_M$ are defined
in terms of the physical Fock states for A's cavity by the \textit{dual rail} basis states
\begin{equation}
\label{0.3}
|{\bf 0}\rangle_M  =  |1\rangle_{A_1}|0\rangle_{A_2},\ \ \ \ |{\bf 1}\rangle_M  =  |0\rangle_{A_1}|1\rangle_{A_2},
\end{equation}
with similar expressions for R's cavity.

The previous construction implicitly assumes that we have chosen a modal decomposition of the Minkowski vacuum
based on intra-cavity and extra-cavity modes. This is a legitimate alternative to the usual way of quantizing the
vacuum in terms of plane wave modes\cite{Dalton}. Note that once the cavities are loaded with a photon, we assume
the cavity is perfect and cannot emit the photon. The quasi modes of reference \cite{Dalton} then become genuine orthogoal modes.

In order to set up a teleportation protocol\cite{NC} we now suppose that Alice has an additional cavity, which
we will call the client cavity (C), again containing  a single qubit with dual rail encoding  by a single photon
excitation of a two mode Minkowski vacuum state. This qubit is in an unknown state
\begin{equation}
\label{0.4}
|\mbf{\psi}\rangle_M =\alpha|{\bf 0}\rangle_M+\beta|{\bf 1}\rangle_M
\end{equation}
As Rob's cavity accelerates away, the client cavity is
brought near to A's cavity so that a joint measurement can be made on the  two orthogonal modes of each cavity.  The joint measurement should correspond to an effective measurement of the two qubit system in the Bell basis for A and C.

The results of this measurement are then sent to Rob, and can be received by him as long as Alice transmits
them before she moves across Rob's horizon (see \Figab{accelframe}{b}). Rob now uses these measurements to make transformations, and
possibly measurements, to verify the protocol  in his local {\em accelerating } frame.  However we now must
confront the possibility that  as Rob is accelerating his cavity will become populated by thermally excited
photons through the Davies-Unruh mechanism \cite{unruh} for accelerated cavities. As we will show, this reduces the fidelity
of a teleportation protocol between accelerated partners.

\subsection{Fidelity of teleported state}
Let us first begin by briefly recalling the usual teleportation protocol, between
Minkowski observers Alice and Bob [\Figab{accelframe}{a}], as given in \cite{NC}.
Our two qubit entangled state will be encoded as entangled Fock states of the electromagnetic field.
Alice wishes to teleport the state $\ketML{\psi} = \alpha \ketML{0} + \beta \ketML{1}$
to Bob. Let Alice and Bob share the entangled Bell state
$\bell{00} = 1/\sqrt{2}(\,\ketML{0}\otimes\ketML{0} + \ketML{1}\otimes\ketML{1}\,)$.
The input state to the system is then $\ketM{\mbf{\Psi}_0} = \ketML{\psi}\,\bell{00}$.
Alice performs a CNOT gate on $\ketML{\psi}$ and her portion of $\bell{00}$,
and then passes the first qubit of the output state through a Hadammard gate.
Upon making a joint projective measurement on her two logical qubits with the result
$\ketML{i}\otimes\ketML{j}$ with $i, j \in \{0,1\}$, the full state is projected
into $\ketML{i}\otimes\ketML{j}\otimes\ketM{\mbf{\phi}_{i,j}}$ where Bob's state is
given by
$\ketM{\mbf{\phi}_{ij}}\equiv x_{ij}\ketML{0} + y_{ij}\ketML{1}$. Here we have defined
the four possible conditional state amplitudes as
$(x_{00},y_{00}) = (\alpha,\beta)$,
$(x_{01},y_{01}) = (\beta,\alpha)$,
$(x_{10},y_{10}) = (\alpha,-\beta)$, and
$(x_{11},y_{11}) = (-\beta,\alpha)$.
After receiving the classical information $\{i,j\}$ of the result of Alice's
measurement, Bob can rotate his qubit of the entangled state into $\ketM{\mbf{\psi}}$
by applying the operations $Z^i_M\,X^j_M$ to $\ketM{{\mbf{\phi}}_{ij}}$, where
$Z$ and $X$ are single qubit rotations on the logical states.
The fidelity of the teleported state is unity in this idealized situation.

Alice now wishes to perform this same teleportation protocol with the
uniformly accelerated Rob.
When Alice sends the result of her measurement $\{i,j\}$, which can
be received by Rob, if Alice has not yet crossed Rob's future horizon $\mathcal{H}_+$,
Rob's state will be projected into (written in the Fock basis)
\bea{6.1}
\lefteqn{\rho^{(I)}_{ij} \equiv \sum_{k=0}^\infty \sum_{l=0}^\infty
\sp _{II}\langle k, l\ketM{\mbf              {\phi}_{ij}} \langle \mbf{\phi}_{ij}\ketII{k,l} }\no
&=& \frac{1}{\cosh^6 r} \Sumnm \Big[  (\tanh^2 r)^{n-1}
                               \left[ (n-m)|x_{ij}|^2  + m |y_{ij}|^2 \right]  \no
& &  \hspace{0.5in} \times \, \ketI{m,n-m}\langle m,n-m| \no
&+& \Big(\, x_{ij} \, y^*_{ij} \tanh^{2n} r \; \sqrt{(m+1)(n-m+1)} \no
& &  \hspace{0.5in} \times \, \ketI{m,n-m+1}\langle m+1,n-m| + \textrm{h.c.}\Big)  \Big].
\eea
In \Eq{6.1} $\ketI{m,n-m} = \ket{m}_{R_1}\otimes\ket{n-m}_{R_2}$ is a state of $n$ total excitations in the region $I$
product state, with $0 \le m \le n$ 
excitations in the leftmost mode.

To obtain $\rho^{(I)}_{ij}$ Rob has expanded the Minkowski Fock states $\{\ketM{0},\, \ketM{1}\}$
appear in the dual rail state $\ketM{\mbf{\phi}_{ij}}$ in terms region $I$ and $II$ Fock states
using \Eq{15} and
\be{6.2}
\ketM{1} = \frac{1}{\cosh^2 r}\, \sum_{n=0}^\infty \, \tanh^n r \,\sqrt{n+1}\,\ketI{n+1}\otimes\ketII{n},
\ee
which results from a simple calculation utilizing \Eq{16} for $\ketM{1} = a^\dag_M \ketM{0}$.
Since Rob is causally disconnected from region $II$, the state he observes must be
adapted to the RRW by tracing out over region $II$.
The state observed by Rob \Eq{6.1}, can be written as
\bea{7.1}
\rho^{(I)}_{ij} &=& \sum_{n=0}^\infty p_n \;\rho^{(I)}_{ij,n} \qquad \textrm{in particular with} \no
\rho^{(I)}_{ij,1} &=& \ketI{\mbf{\phi}_{ij}} \langle \mbf{\phi}_{ij}|,  \qquad p_0 = 0, \;  p_1 = 1/\cosh^{6}r .
\eea
The block tridiagonal form of $\rho^{(I)}_{ij}$ is illustrated in \Fig{densitymatrix}.

\begin{figure}[h]
\centering
$$
\hspace*{-2em}
\begin{array}{cc}
\mbox{} &
\hspace{-.5em}\begin{array}{llllll}
  \hspace*{-1em} \ketI{00}   & \hspace*{1em}\ketI{01} \hspace*{0em} & \ketI{10} \hspace*{1em}&
  \ketI{02} \hspace*{1.5em}   & \ketI{11} \hspace*{1em} & \hspace*{1em} \ketI{20}  \\
\end{array} \\
\begin{array}{c}
  \braI{00} \\ \braI{01} \\ \braI{10} \\ \braI{02} \\ \braI{11} \\ \braI{20}  \\
\end{array} &
\hspace{-.5em}\left(%
\begin{array}{cccccc}
  \hspace*{1em} 0 & \mbox{} & \mbox{}  & \mbox{}  & \mbox{}  & \mbox{}  \\
  \mbox{}  & \hspace*{1em}|x_{ij}|^2 & x_{ij} y^*_{ij} & \mbox{}  & \mbox{} & \mbox{}  \\
  \mbox{}  & \hspace*{1em} x^*_{ij} y_{ij} & |y_{ij}|^2 & \mbox{}  & \mbox{}  & \mbox{}  \\
  \mbox{}  & \mbox{}  & \mbox{}  & 2|x_{ij}|^2 & \sqrt{2}\, x_{ij} y^*_{ij} & 0 \\
  \mbox{}  & \mbox{}  & \mbox{}  & \sqrt{2}\, x^*_{ij} y_{ij} & 1 & \sqrt{2} \,x_{ij} y^*_{ij} \\
  \mbox{}  & \mbox{}  & \mbox{}  & 0 & \sqrt{2}\, x^*_{ij} y_{ij} & 2|y_{ij}|^2 \\
\end{array}%
\right) \\
\end{array}
$$
\caption{A schematic representation of $\rho^{(I)}_{ij}$ of \Eq{6.1} illustrating the
$n=\{0,1,2\}$ excitation sectors. For the $0$-excitation sector $\rho^{(I)}_{ij,0}$ has
probability $p_0\equiv 0$,
since even at zero acceleration there is at least one excitation in the system
from the initial entanglement creation. The $1$-excitation sector spanned
by $\{\ketI{01},\ketI{10}\}$ has probability $p_1 = 1/\cosh^{6} r$ and is given by
$\rho^{(I)}_{ij,1} = \ketI{\mbf{\phi}_{ij}} \langle \mbf{\phi}_{ij}|$, which is a pure state.
The $n\ge 2$-excitation sectors $\rho^{(I)}_{ij,n}$ arise from the presence of Davies-Unruh
radiation. They have probability $p_n = (\tanh^2 r)^{n-1}/\cosh^{6} r$
and are mixed states; $n=2$ is illustrated.
}
\label{densitymatrix}
\end{figure}

Upon receiving the result $(i,j)$ of Alice's measurement, Rob can apply the rotation operators
$\left.Z^i_I\,X^j_I\right|_{n=1}$ restricted to the 1-excitation sector of his state spanned by
$\{\ketIL{0}, \ketIL{1}\} = \{\ket{1}_{R_1}\otimes\ket{0}_{R_2}, \ket{0}_{R_1}\otimes\ket{1}_{R_2}\}$ to turn this portion of
his density matrix into the region $I$ analogue of the state Alice attempted to teleport to him, namely
\be{8.1}
\ketI{\mbf{\psi}} =  \, \alpha \ketIL{0} + \beta \ketIL{1} .
\ee
The fidelity of Rob's final state with $\ketIL{\psi}$ is then
\be{9.1}
F^{(I)} \equiv Tr_I\Big(\ketI{\mbf{\psi}}\langle\mbf{\psi}|\,\rho^{(I)}\Big)
= \sp_I\langle\mbf{\psi}|\rho^{(I)}\ketI{\mbf{\psi}} = 1/\cosh^6 r.
\ee

At $r=0$, corresponding to $a=0$, we
are back to the usual case of teleportation between Alice and Bob, both Minkowski observers,
and the fidelity is unity. Using the definition of $\Omega$ and $r$ in
\Eq{5b} we find
\bea{22.1}
\tanh r &=&  \exp[-\pi\omega_R / (a / c)] \quad \textrm{or} \no
r &\approx& \exp[-\pi\omega_R / (a / c)] \quad \textrm{for} \quad r\to 0.
\eea
For terrestrial experiments where $a=g\sim 10 \, m/s^2$, $a/c \sim 10^{-8} \,s^{-1}$
is such a small frequency, that for all frequencies $\omega_R$ of physical interest
the fidelity is near unity with incredible precision, (for $r=10^{-3}$ the fidelity
is is unity to within one part in $10^6$, and this still corresponds to
unphysically large accelerations).
Near the event horizon of a black hole appreciable accelerations can be obtained
such that the reduction of the fidelity from unity could be observed.
In that case an analogous teleportation scheme could be defined with the accelerated Rob stationary
outside the event horizon, and Alice freely falling into the black hole.

Rob can check to see whether any
Davies-Unruh photons have been excited in his local cavity using a non-absorbing detector. The probability
of obtaining the  answer NO is $\cosh^{-6}r$.
In that case, he has restored the required
entanglement, and teleportation can be completed without error.
However, if Rob obtains the more likely result YES, then he cannot complete the protocol without error. In principle
Rob may be able implement an error correction protocol that simply looks for more than one photon in
each of his modes and makes the appropriate correction.

\subsection{Reduction of Fidelity in terms of Entropy}
It is of some interest to consider the reduction of fidelity  in terms of entropy.
Consider the the von Neumann entropy $S = - Tr(\rho^{(I)} \log \rho^{(I)})$
of Rob's pre-measurement state, post-measurement state upon learning the result
of Alice's measurement, and the vacuum state, as a function of $r$.
The pre-measurement state is obtained from \Eq{6.1} by averaging $(i,j)$ over all four possible input states,
which reduces it to a diagonal density matrix.
The post-measurement state is given by \Eq{6.1} with the input state to the teleportation
protocol chosen to be $\ketML{\psi} = 1/\sqrt{2} \, \big(\ketML{0} + \ketML{1}\big)$,
without loss of generality.

For any acceleration $r(a)$, the 1-excitation sector of $\rho^{(I)}$ is always
$\big(\ketI{\mbf{\phi}_{ij}} \langle \mbf{\phi}_{ij}|\big)/\cosh^6 r$. For the particular
choice of $x_{ij} = y_{ij} = 1/\sqrt{2}$ for the teleported state, the probabilities
of Rob's diagonal pre-measurement state are given by
$p^{pre}_{n,m} = n/2 \, (1-\xi)^3 \, \xi^{n-1}\equiv p^{pre}_{n}$, independent of $m$ for
$n\ge 0$ and $0\le m \le n$,
with $\xi\equiv \tanh^2 r$. The eigenvalues of the post-measurement state
are given by $p^{post}_{n,m} = m \, (1-\xi)^3 \, \xi^{n-1}$ for $n\ge 1$ and $0\le m \le n$,
with $p^{post}_{0,0}\equiv 0$.
As the acceleration increases to infinity (i.e. $r\to\infty$; $\xi\to 1$),
the higher $n$-excitation density matrices $\rho^{(I)}_{ij,n}$ of \Eq{6.1} make
their presence known with probability proportional to $(1-\xi)^3 \, \xi^{n-1}$.
The relationship between eigenvalues of Rob's pre-measurement state, before
he receives the result of Alice's measurement, and his post-measurement state is
$p^{pre}_{n} = (n+1)^{-1} \sum_{m=0}^n p^{post}_{n,m}$ where $n+1$ is the number of
states of the form $\ketI{m,n-m}$ for fixed $n$, spanning $\rho^{(I)}_{ij,n}$.
\begin{figure}[h]
\centering
\includegraphics[height=2.5in,width=2.5in]{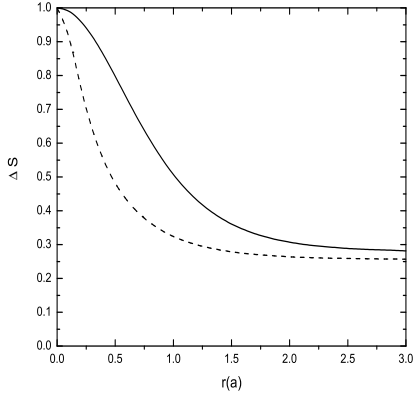}
\caption{Rob's entropy information gain (in bits) $\Delta S_{gain}
= S_{pre} - S_{post}$ upon receiving the Alice's measurement
results: numerical (solid) and $\Delta S_{gain}^{n=2}$ (dashed) for a 5-state,
$n=\{1,2\}$ excitation model using the
$\{\ketI{0\,1},\ketI{1\,0}\}$ and $\{\ketI{0\,2},\ketI{1\,1}, \ketI{2\,0}\}$
sector from Rob's post measurement state, with $x_{ij} = y_{ij} = 1/\sqrt{2}$.
Here $\ketI{k\,l} = \ket{k}_{R_1}\otimes\ket{l}_{R_2}$.}
\label{dSvsr}
\end{figure}
It is worthwhile to note that the  Minkowski vacuum state Rob moves
through is perceived by him as the thermal Rindler state $\rho^{(I)}_{\vacM} \equiv Tr_{II}(\vacM\langle 0|)$
with diagonal entries $p^{vac}_{n,m} = (1-\xi)^2 \, \xi^{n} \equiv p^{vac}_{n}$ for
$n\ge 0$ and $0\le m\le n$. In a sense, each normalized $n$-excitation
density matrix ot the pre-measurement state is individually thermalized with
equal entries proportional to $\xi^{n-1}$ as opposed to $\xi^n$ as in $\rho^{(I)}_{\vacM}$.
Within the same $n$-excitation subspace, the post-measurement state retains a character
distinct from a thermalized state, with probabilities proportional to $m x^{n-1}$
for each of its $n+1$ diagonalized component states.

The difference of the von Neumann entropies
between $\rho^{(I)}_{pre}$ and $\rho^{(I)}_{post}$ is plotted in \Fig{dSvsr}
along with a normalized $5$-state model incorporating the $n=\{1,2\}$ excitation sectors of
both density matrices. Individually, the components of each density matrix
are approaching zero due to the factors of $(1-\xi)^3 \, \xi^{n-1}$.
However, the observation that both the complete and approximate
model show that $\Delta S \equiv S_{pre} - S_{post}$ approaches zero
very slowly (note that $r=3$ in \Fig{dSvsr} implies $\xi\sim 0.58$) indicates
that $\rho^{(I)}_{post}$ retains a non-thermalized nature in each $n$-excitation
subspace for finite $r$. As a whole, Rob's state $\rho^{(I)}_{post}$ is being thermalized by his acceleration
through the Minkowski vacuum, which he perceives as a thermal state,
and asymptotically $\lim_{r\to\infty (x\to 1)} \Delta S = 0$.


\section{Teleportation with Dirac Particles}
In this section we extend the previous results for polarizationless entangled photon states,
modelled by massless modes of a bosonic scalar field, to that for entangled spin $1/2$ Dirac states,
which we can consider as electrons. We follow the same accelerated partner teleportation
protocol as above and focus on the changes that arise from the switch from bosonic to
fermionic particles. In the end, the resulting degradation in the fidelity of the
teleported state and the reduction in Rob's information gain as a result of his
acceleration is analogous to that for the previous case involving bosons. The specific
differences arise from the crucial sign change in Fermi-Dirac versus the Bose-Einstein
distribution, and the finite number of allowed excitations allowed in fermionic
systems due to the Pauli exclusion principle over the unbounded excitations that
can occur for the bosonic harmonic oscillator.

The study of spin fields in a uniformly accelerated frame was first carried out
by Candelas and Deutsch \cite{candelas}. The Dirac field in a Rindler frame
was latter investigated by many authors \cite{greiner,hacyan,bautista,mastas}. One important
result for our purposes that came out of these studies was that the spinor
for the accelerated observer can be obtained from the unaccelerated Minkowski
observer by a linear transformation (composed of projection operators of
the form $\mbf{P}_\pm = 1/2(1\pm \gamma^0\gamma^3)$, where $\gamma^\mu$ are the Minkowski $4\times 4$ Dirac matrices
satisfying $[\gamma^\mu, \gamma^\nu] = 2 \,\eta^{\mu\nu}\, \mbf{1}_{4\times 4}\,$ \cite{hacyan})
which does not mix spin components. Thus, for example, a spin up Minkowski state remains
a spin up state if it undergoes uniform acceleration. In fact, for the Dirac equation in a general gravitational
field, it can be shown \cite{QMPhaseGR} that for a constant amplitude (though momentum dependent) WKB solution
of the form $\psi = a \exp(i S(x)/\hbar)$ the gravitational field does not affect the spin.
In the next order of approximation, where $a\to a(x)$, one finds that the spin is parallel propagated
along the path of the Dirac particle \cite{lammerzahl} (and references therein)  i.e. the spin behaves in the same way
as a spinning top. For gravitational theories with torsion, the spin couples to the torsion, while the
orbital angular momentum does not. It is only in the next higher order of the WKB approximation that one
finds a dependence of the spin on the path of the wave packet (i.e. the spin creating non-geodesic motion)
\cite{audretsch2}.  There is a slight technical difference in the definition of the spin operator between
the work of \cite{hacyan} and \cite{bautista} that arises for the Dirac equation in a Rindler
spactime when an external electromagnetic field is applied. In the absence of an externally
applied field, as considered in this work, the different definitions reduce to an identical form.
In the following, we draw upon results from the paper by Hacyan \cite{hacyan}
and adapt them to our analysis.

In the single mode analysis that we consider in this work, the bosonic Minkowski vacuum \Eq{15}
can be written as $\ketM{0} = \mathcal{N}_b\exp(\tanh r \,b^\dag_{I}\,b^\dag_{II})\,\ketI{0}\otimes\ketII{0}$,
where $\mathcal{N}_b = 1/\cosh r$ is the bosonic normalization factor and $\tanh r = \exp(-\pi\Omega)$,
with $\Omega \equiv \omega_R / (a/c)$.
For the case of fermions we in essence have an complex rotation of $r\to i\,r$ which yields
$\ketM{0} = \mathcal{N}_f\exp(i\tan r \,c^\dag_{I}\,c^\dag_{II})\,\ketI{0}\otimes\ketII{0}$,
where $c^\dag_I,\,c_I$ and $c^\dag_{II},\,c_{II}$ are fermionic creation and annihilation
operators in region $I$ and region $II$, $\mathcal{N}_f = \cos r$ is the fermionic normalization factor,
and $r$ is now defined through $\tan r = \exp(-\pi\Omega)$. In analogy with \Eq{5b}
the fermionic Bogoliubov coefficients \cite{greiner,hacyan} allow us to define
\be{17}
\cos r = \tOmega, \; \sin r = \expmpiOmega \, \tOmega,
\ee
where $r$ is restricted to the range $0\le r(a) \le \pi/4$ corresponding to
$0\le a \le \infty$. The resulting crucial sign change in \Eq{17} traces back
to the use of anti-commutation relations for fermionic operators versus
the commutation relations for bosonic operators. Absorbing factors of $i$ into the definition
of the Fock states, the single mode fermionic Minkowski vacuum can be written as (compare \Eq{15})
\be{18}
\ketM{0} = \cos r \ketI{0}\otimes\ketII{0} + \sin r \ketI{1}\otimes\ketII{1},
\ee
in accordance with the Pauli exclusion principle which limits the number of fermionic
excitations in any Fock state to $\{0,1\}$.
The single mode fermionic vacuum perceived by Rob is given by
\be{18.1}
\rho^{(I)}_{\vacM} \equiv Tr_{II}(\vacM\langle 0|) =
\cos^2 r\, \ketI{0}\langle 0| + \sin^2 r\, \ketI{1}\langle 1| .
\ee
In this single mode
approximation, a Minkowski creation operator can be written as (compare \Eq{16})
\be{18.2}
a^\dag_M = \cos r \, c^\dag_I + \sin r \, c_{II}.
\ee
Using \Eq{18.2} it is simple to show that the $1$-excitation Minkowski Fock state is given by
\be{19}
\ketM{1} = a^\dag_M \, \ketM{0} = \ket{1}_I \otimes \ketII{0},
\ee
and that $(a^\dag_M)^2 \, \ketM{0} \equiv 0$, preserving the Pauli exclusion principle.

We are now in a position to repeat the accelerated partner teleportation protocol of
the previous section. We interpret dual rail basis states $\{\ketML{0}\, \ketML{1} \}$
appearing in the Bell state \Eq{0.2} as an excitation of a spin up state in
one of two possible spatial modes in Alice's cavity (and similarly for Rob) as in \Eq{0.3}.
The calculation proceeds straight forwardly as before, with even simpler algebra.
When Alice makes the joint measurement on her two states with result $\{i,j\}$,
Rob state is projected into (compare \Eq{6.1})
\bea{20}
\rho^{(I)}_{ij} &\equiv& \sum_{k=0}^1 \sum_{l=0}^1
\sp _{II}\langle k\,l\ketM{\mbf{\phi}_{ij}} \langle \mbf{\phi}_{ij}\ketII{k\,l} \no
&=& \cos^2 r \, \ketI{\mbf{\phi}_{ij}} \langle \mbf{\phi}_{ij}|
+  \sin^2 r \, \ketI{11}\langle 11 |.
\eea
In particular, $\ketI{11}\equiv \ket{1}_{R_1}\otimes\ket{1}_{R_2}$
is a state of two excitations, one in each of the two spatial modes of Rob's cavity.
In the bosonic case, a state of two excitations would also include the
basis states $\ket{0}_{R_1}\otimes\ket{2}_{R_2}$ and $\ket{2}_{R_1}\otimes\ket{0}_{R_2}$ (see \Fig{densitymatrix}).
However, from the Pauli exclusion principle, these and all other states of excitations
$n\ge 2$ are excluded.

As in the previous section, upon receiving the result $(i,j)$ of Alice's measurement, Rob can apply the rotation operators
$\left.Z^i_I\,X^j_I\right|_{n=1}$ restricted to the 1-excitation sector of his state spanned by
$\{\ketIL{0}, \ketIL{1}\} = \{\ket{1}_{R_1}\otimes\ket{0}_{R_2}, \ket{0}_{R_1}\otimes\ket{1}_{R_2}\}$ to turn this portion of
his density matrix into the region $I$ analogue of the state Alice attempted to teleport to him, namely
\be{21}
\ketI{\mbf{\psi}} =  \, \alpha \ketIL{0} + \beta \ketIL{1} .
\ee
The fidelity of Rob's final state with $\ketIL{\psi}$ is then
\be{22}
F^{(I)} \equiv Tr_I\Big(\ketI{\mbf{\psi}}\langle\mbf{\psi}|\,\rho^{(I)}\Big)
= \sp_I\langle\mbf{\psi}|\rho^{(I)}\ketI{\mbf{\psi}} = \cos^2 r.
\ee
For the bosonic case \Eq{9.1} the fidelity $1/\cosh^6 r$ decreases to zero
as the acceleration approaches infinity. This occurs since the thermal excitations
due to the Davies-Unruh radiation can run up the infinite excitation ladder
accommodated by the harmonic oscillator, while the desired teleported state $\ketIL{\psi}$ resides in
the $1$-excitation sector. In the fermionic case, the number of allowed excitations
is bounded above by $n=2$, again with the desired teleported state residing in the $1$-excitation sector.
Thus as the acceleration approaches infinity, the fidelity saturates to $\lim_{r\to\pi/4} \cos^2 r = 1/2$.

From an entropy point of view, Rob's pre-measurement state is given by
\be{23}
\rho_{pre}^{(I)} = \frac{1}{2}\cos^2 r \big( \ketI{0\,1}\langle 0\,1| + \ketI{1\,0}\langle 1\,0| \big)
+ \sin^2 r \ketI{1\,1}\langle 1\,1|,
\ee
and the post measurement state, again for the choice of $x_{ij} = y_{ij} = 1/\sqrt{2}$, is given by
\be{24}
\rho_{post}^{(I)} = \frac{1}{2}\cos^2 r \, \ketIL{\psi}\langle \mbf{\psi} | +  \sin^2 r \, \ketI{11}\langle 11 |.
\ee
This latter state is easily diagonalized and it is found that the von Neumann entropies (in bits)
satisfy the relation $S_{pre} = S_{post} + \cos^2 r$ and hence Rob's information gain is
given by
\be{25}
\Delta S_{gain} = S_{pre} - S_{post} = \cos^2 r .
\ee
At zero acceleration, $r=0$, with Alice and Rob both Minkowski,
$\Delta S_{gain} = 1$ states that Rob can completely restore the
properties of the state that Alice teleports to him. At infinite
acceleration, $r=\pi/4$, Rob's information gain saturates at a value
$\Delta S_{gain}=1/2$ again due to the finite number of excitations
allowed by the fermionic system.


\section{Discussion and Conclusions}
The main issues of teleportation between an inertial Minkowksi observer Alice
and a non-inertial, uniformly accelerated Rindler observer Rob are two fold.
First, as a result of the acceleration, the Minkowski vacuum that Rob moves
through (for a single Rindler mode $(\Omega,\kp)$) can be written as a
two-mode squeezed state with the component Fock states existing in causally
disconnected regions $I$ and $II$. Second, as a result of this fact, Rob perceives the
Minkowski vacuum as a pure thermal state of temperature
$T_U$ as the inevitable result of his complete ignorance of region $II$.
In an attempt to teleport a state $\ketM{\mbf{\psi}} = \alpha \ketML{0} + \beta \ketML{1}$
to Rob, the best we can expect this uniformly accelerated observer to recover at the end of the protocol is
$\ketI{\mbf{\psi}} = \alpha \ketIL{0} + \beta \ketIL{1}$. We have shown that
the fidelity of Rob's post-measurement state with this best possible result $\ketIL{\psi}$ is $\cosh^{-6} r$.
In addition, we have demonstrated that the information gain obtained by Rob (defined
as the difference in the von Neumann entropies of his pre- and post-measurement states),
decreases with increasing acceleration through the Minkowski vacuum, which Rob
perceives as a Rindler thermal state.
At high acceleration (high Davies-Unruh temperatures) all information
is lost and Rob perceives only the thermalized vacuum state.

Recently,  Anderson \textit{et al} \cite{vanEnk} have also
discussed  teleportation and the Unruh vacuum. However, that work
considers a situation that is physically quite different
than the one presented here.
The authors use the \textit{mirror modes} of Audretsch and
M\"{u}ller \cite{audretsch} and consequently have the accelerated
observers travelling on oppositely directed hyperbolas, with Alice
in region $I$ and Bob in the causally disconnected region $II$.
The teleportation protocol is then interpreted from the point of
view of a Minkowski observer Mork. In this work, we consider a
setup between observers, one stationary, the other accelerated,
who remain causally connected to each other during the
teleportation protocol.
Kok and Yurtsever \cite{ulvi} have recently considered the
interaction of a uniformly accelerated qubit with a massless
scalar field (in a similar fashion to the classic "particle detector"
calculation of Unruh and Wald \cite{UnruhWald}) and show that
the qubit decoheres. For long interaction times and slow enough
accelerations the decoherence can be made arbitrarily small.
In general, a more realistic, though much more difficult analysis
would take into account any decoherence effects arising from Rob's motion from
zero to some finite constant acceleration.
B.L. Hu \textit{et al} \cite{blhu}
have also considered the motion of an arbitrary number of detectors modelled as
oscillators and minimally coupled to a massless scalar field in $1+1$ dimensions.
In this approach, the scalar field is integrated out, leaving a reduced
set of effective semiclassical equations which nonetheless contain the full
quantum dynamics of the field. For an accelerated detector and probe the main
contributions arise from field correlations across the horizon.

The model investigated here is analogous to teleportation through two channels,
one of which is free space for Alice and the second which involves parametric
down conversion with the following caveat described below. In the second channel, a signal
mode $I$ and an idler mode $II$ experience a squeezing Bogoliubov transformation analogous to
\Eq{14} \cite{yurke} (see appendix \ref{appendix_vacuum_state}).
Here $r$ is proportional to the coupling strength between
the signal and idler mode times the length of the crystal through which
the parametric down conversion takes place; higher interaction strengths and/or
longer interaction lengths corresponds to a higher Davies-Unruh temperature.
The caveat is that Rob, acting as say the signal mode, has no access
to information about the idler and therefore must trace out this information.
Performing the teleportation protocol in such a system is analogous to
teleportation between a Minkowski and Rindler observer as considered in this work.
In the parametric down conversion model, Rob can choose to ignore the idler
information thus mimicking a Rindler observer. However, for an
accelerated observer, the existence of the horizons $\mathcal{H}_\pm$ is of fundamental
importance. Since region $I$ and $II$ are causally disconnected, there is
no way, even in principle, for Rob to have any information about region $II$,
and thus his state is always a reduced density matrix appropriate for region $I$.

The work presented here also has implications for teleportation in curved
spacetimes.  Recently, the
equivalence of Hawking and Davies-Unruh temperatures has been established by
embedding curved spacetimes in flat spaces of sufficiently high
dimension, and then computing the Davies-Unruh temperatures of uniformly accelerated
observers in the latter \cite{deser}. For a scalar field near a horizon,
the wave equation takes a universal free field form, and the mode decomposition
is essentially the same as in the Rindler case. The analysis for curved spacetimes
would proceed in an analogous fashion as the one presented here, but
in the higher dimensional flat space, and teleportation would again be
degraded by the observer's acceleration. Details will be expanded upon
in future work.

We have given an explanation of the reduction of teleportation
fidelity in terms of the Davies-Unruh radiation seen by Rob in his frame.
Note that this is an operationally meaningful statement as Rob can
attempt to verify that he has received the desired state
$\big(x_{lm}|\mbf{0}\rangle_I + y_{lm}|\mbf{1}\rangle_I\big)$ by local verification
measurements (e.g. a single photon interference experiment in the bosonic case), and then
send the results to Alice.   From an operational point of view Alice
would conclude that the shared entangled  resource  has become
decohered. It is well know that entanglement is a fragile resource
in the presence of environmental decoherence. It appears also to be a fragile
resource when one of the entangled parties undergoes acceleration.
While the degree of decoherence  is exceedingly small for practical
accelerations, the apparent connection between space time geometry
and quantum entanglement is intriguing.


\appendix
\section{Origin of the Two-Mode Squeezed State Minkowski Vacuum and the Thermal Rindler Vacuum}
\label{appendix_vacuum_state}
The origin of the bosonic two-mode squeezed state vacuum given in \Eq{4} and \Eq{15} and
the subsequent thermal Rindler vacuum given by \Eq{7} is a simple
consequence of the commutation relations of the Minkowski
and Rindler creation and annihilation operators \Eq{16}. Following Yurke \cite{yurke}
let us consider the analogous process of  parametric down conversion between a signal
and idler mode with annihilation operators $a_S$ and $a_I$ respectively.
Here we treat the pump as non-depleted. For a vacuum state input (analogous to
the Minkowski vacuum $\ketM{0}$), the operators $b_S, b_I$ given at the output of the squeezing crystal
have the general form
\bea{27}
b_S &=&  S_{11} a_S + S_{12} a^\dagger_I \no
b^\dagger_I &=&  S_{21} a_S + S_{22} a^\dagger_I .
\eea
The commutation relations $[b_S,b_S^\dagger]=1$, $[b_I,b_I^\dagger]=1$ and $[b_S,b_I]=0$
result in the constraints
\bea{28}
|S_{11}|^2 - |S_{12}|^2 &=& 1 \no
|S_{22}|^2 - |S_{21}|^2 &=& 1 \no
S_{11} S^*_{21} &=& S_{12} S^*_{22}.
\eea
These relationships can be inverted to write the input operators $a$ in terms of the
output operators $b$ as
\bea{29}
a_S &=&  S^*_{11} b_S - S^*_{21} b^\dagger_I \no
a^\dagger_I &=&  -S^*_{12} b_S + S^*_{22} b^\dagger_I .
\eea
The hyperbolic relations in \Eq{28} allow us to write
\bea{29b}
S_{11} = \cosh r, &\qquad& S_{12} = e^{i\phi} \sinh r \no
S_{21} = e^{i\phi} \sinh r, &\qquad& S_{22} =  \cosh r,
\eea
as used in \Eq{16} (with the unimportant phase $e^{i\phi}$ absorbed into the definition of the operators).
For the case of parametric down conversion,
$r$ is proportional to the interaction strength times the length
of the crystal. The above relationships also imply
\be{30}
b^\dagger_S b_S - b^\dagger_I b_I = a^\dagger_S a_S - a^\dagger_I a_I.
\ee
Thus, the difference in the number of signal and idler photons leaving
the squeezing crystal is the same as the initial
signal and idler difference at input. In particular, if the input is
the vacuum, \Eq{30} requires that the photons must exit the crystal in pairs,
 each signal photon matched with a corresponding idler photon.

The  signal and idler output vacuum states $|0\rangle_S$ and $|0\rangle_I$
defined such that $b_S |0\rangle_S = 0$ and $b_I |0\rangle_I = 0$
(analogous to the Rinder vacuums in region $I$ and $II$) are different
than the vacuum $|0\rangle_a$ (analogous to the Minkowski vacuum $\ketM{0}$)
annihilated by the input operators $a_S$ and $a_I$. Since the signal and idler
photons are created in pairs the form of the two-mode squeezed vacuum $|0\rangle_a$
can be taken to be
\be{31}
|0\rangle_a = \sum_{n=0}^{\infty} A_n | n \rangle_S \otimes | n \rangle_I.
\ee
From the requirement that $a_S |0\rangle_a = 0 = a_I |0\rangle_a$, and using \Eq{29}
we have the equation $(S^*_{11} b_S - S^*_{21} b^\dagger_I )|0\rangle_a = 0$ with solution
$A_{n+1} = (S^*_{21}/S^*_{11})\, A_n$ or $A_n = (S^*_{21}/S^*_{11})^n A_0$. Normalization
of the vacuum state $_a\langle 0 | 0 \rangle_a = 1$ produces $|A_0|^2 = |S_{11}|^{-2}$ and yields the state
\bea{32}
| 0 \rangle_a &=& \frac{1}{|S_{11}|} \sum_{n=0}^{\infty}
\left( \frac{S^*_{21}}{S^*_{11}}\right)^n | n \rangle_S \otimes | n \rangle_I \no
&=& \frac{1}{\cosh r} \sum_{n=0}^{\infty} \tanh^{n} r \;| n \rangle_S \otimes | n \rangle_I,
\eea
where in the last step, \Eq{29b} was used. Equation(\ref{32}) is just \Eq{15}.

If one of the output modes is unobserved, the state describing the observed mode
is a thermal vacuum state. This follows from the arguments in Section II where
for, say, an un-observed signal mode, we take the observed idler state as
\be{33}
\rho^{(I)}_{|0\rangle_a} \equiv Tr_S(|0\rangle_a\langle 0|) =
\left(
1-\expm2piOmega  \right) \Sum \expnOmega |n\rangle_I \langle n |.
\ee
In \Eq{33} we have made the identifications as in \Eq{5b}
\bea{34}
 \cosh r &=& \sOmega,  \no
 \sinh r &=& \expmpiOmega \, \sOmega, \quad \Omega \equiv \frac{\omega_I}{a/c},
\eea
where $\omega_I$ is the frequency of an idler photon.
In general the Davies-Unruh temperature $T_U$ is given by
\be{35}
T_U = \frac{\hbar\omega_I}{2 \pi k_B \ln(|S_{11}|/|S_{21}|)}
\ee
which reduces to \Eq{8} in the case of an accelerated Rindler observer
with $a \to \omega_I c / \ln(|S_{11}|/|S_{21}|)$.

\acknowledgements
The authors wish to thank Jonathan P. Dowling, Ulvi Yurtsever and the members of the
JPL Quantum Computing Technologies Group for stimulating discussions
on this and other topics in relativistic quantum information theory.
GJM acknowledges useful discussions with Paul Davies.


\end{document}